\documentclass{imamat}

\usepackage{latexsym,amssymb,lastpage}
\usepackage{graphicx,amsfonts}
\usepackage{times,mathptmx,bm,amsmath}
\usepackage{dcolumn}
\usepackage{graphicx,amssymb,bm}
\usepackage{mathrsfs}
\usepackage{amssymb}
\usepackage{amsmath}
\usepackage{tikz}
\usepackage{color}

\usepackage{float,placeins}

\usetikzlibrary{decorations.markings}
\usetikzlibrary{decorations.pathmorphing}
\usetikzlibrary{fit}
\usetikzlibrary{calc,arrows,patterns}
\usepgflibrary{plotmarks}

\newcommand{\with}{\mathrm{with}}

\newcommand{\tr}{\mathrm{tr}\,}

\newcommand{\where}{\mathrm{where}}

\newcommand{\for}{\mathrm{for}}

\newcommand{\chain}{\mathrm{chain}}

\newcommand{\oi}{\mathrm{ortho}}
\newcommand{\iso}{\mathrm{iso}}
\newcommand{\T}{\mathrm{T}}

\newcommand{\tanhs}{\mathrm{\tanh\,}}
\newcommand{\C}{\mathrm{cyc}}
\newcommand{\s}{\phantom{0}}

\numberwithin{equation}{section}

\begin{document}

\title{Orthotropic cyclic stress-softening model for pure shear during repeated loading and unloading
}

\author{{\sc Stephen R. Rickaby}\footnote{Email: stephen.r.rickaby@gmail.com} 
{\sc and} {\sc Nigel H. Scott}
\footnote{Email: n.scott@uea.ac.uk}\\[2pt]
{School of Mathematics, University of East Anglia,\\[2pt] Norwich Research Park,
 Norwich NR4 7TJ, UK}\\[6pt]
{\rm [Received on 25 December 2013; accepted on 15 April 2014; Online on 28 May 2014]}}

\maketitle
\pagestyle{headings}
\markboth{{\sc S. R. Rickaby and N. H. Scott}}{{\sc Orthotropic Cyclic Stress-Softening}}

\begin{center}
\emph{Dedicated to Ray Ogden on the occasion of his 70th birthday}
\end{center}

\begin{abstract}
{We derive an orthotropic model to describe the cyclic stress-softening of a carbon-filled rubber vulcanizate through multiple stress-strain cycles with increasing values of the maximum strain. We specialize the deformation to pure shear loading.  As a result of strain-induced anisotropy following on from initial primary loading, the material may subsequently be described as orthotropic because in pure shear there are three different principal stretches so that the strain-induced  anisotropy of the stress response  is  different in each of these three directions. We derive non-linear orthotropic models for the elastic response, stress relaxation and residual strain in order to model accurately the inelastic features associated with cyclic stress softening.  We then develop an orthotropic version of the Arruda-Boyce eight-chain model of elasticity and then combine it with the ideas previously developed in this paper to produce an orthotropic constitutive relation for the cyclic stress-softening of a carbon-filled rubber vulcanizate. The  model developed here includes the widely-occurring effects of hysteresis, stress-relaxation and residual strain. The model is found to compare  well with experimental data. }
{Mullins effect, stress-relaxation, hysteresis, residual strain, orthotropy.}
\end{abstract}

\section{Introduction}     

When a rubber specimen is loaded, unloaded and then reloaded, the subsequent load required to produce the same deformation is smaller than that required during primary loading. This stress-softening phenomenon is known as the Mullins effect,  named after Mullins \cite{mullins1947} who conducted an extensive study  into carbon-filled rubber vulcanizates. Diani \textit{et al.} \cite{dianib} have written a   review of this effect, detailing specific features associated with stress-softening and providing a pr\'ecis of models developed to represent this effect.

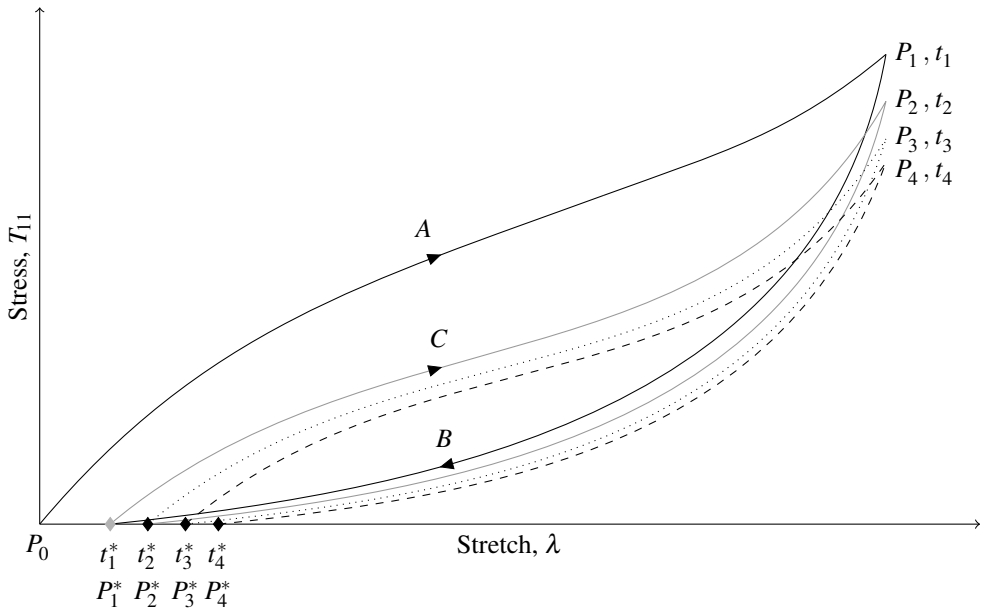
\begin{figure}[H] 
\centering
\begin{tikzpicture}[scale=1.25, decoration={
markings,
mark=at position 6.5cm with {\arrow[black]{triangle 45};},
mark=at position 17.5cm with {\arrowreversed[black]{triangle 45};},
mark=at position 43cm with {\arrow[black]{triangle 45};},}
]
\draw[->] (0,0) -- (10,0)node[sloped,below,midway] {Stretch, $\lambda$} ;
\draw[->] (0,0) -- (0,5.5) node[sloped,above,midway] {Stress, $T_{11}$};
\draw [black](0,0) to[out=50,in=200] node [sloped,above] {} (6,3.5);
\draw [black](6,3.5) to[out=20,in=220] node [sloped,above] {} (9,5);
\draw [black] (0.75,0) to [out=6,in=260] node [sloped,above] {} (9,5);
\draw [black!40] 
(0.75,0) to[out=40,in=240] node [sloped,above] {} (9,4.5)
(1.15,0) to[out=6,in=256.5] node [sloped,below] {} (9,4.5);
\draw [dotted]
(1.15,0) to[out=40,in=238] node [sloped,below] {} (9,4.1)
(1.55,0) to[out=6,in=253] node [sloped,above] {} (9,4.1);
\draw [dashed] 
(1.55,0) to[out=40,in=236] node [sloped,below] {} (9,3.85)
(1.9,0) to[out=6,in=250] node [sloped,above] {} (9,3.85);
\draw [postaction={decorate}][loosely dotted,line width=0.01pt,black!40]
(0,0) to[out=50,in=200] node [sloped,above] {} (6,3.5)
(6,3.5) to[out=20,in=220] node [sloped,above] {} (9,5)
	(0.75,0) to[out=6,in=260] node [sloped,above] {} (9,5)
	(0.75,0) to[out=40,in=240] node [sloped,above] {} (9,4.5)
	(0.75,0) to[out=40,in=240] node [sloped,above] {} (9,4.5);
\coordinate [label=right:{$P_1^{\phantom{*}}, \,t_1^{\phantom{*}}$}] (B) at (9.0	,5);
\coordinate [label=right:{$P_2^{\phantom{*}}, \,t_2^{\phantom{*}}$}] (B) at (9.0	,4.5);
\coordinate [label=right:{$P_3^{\phantom{*}}, \,t_3^{\phantom{*}}$}] (B) at (9.0	,4.1);
\coordinate [label=right:{$P_4^{\phantom{*}}, \,t_4^{\phantom{*}}$}] (B) at (9.0	,3.75);
\coordinate [label=below:{$t_1^*$}] (B) at (0.75,-0.1);
\coordinate [label=below:{$t_2^*$}] (B) at (1.15,-0.1);
\coordinate [label=below:{$t_3^*$}] (B) at (1.55,-0.1);
\coordinate [label=below:{$t_4^*$}] (B) at (1.9,-0.1);
\coordinate [label=below:{$P_1^*$}] (B) at (0.75,-0.5);
\coordinate [label=below:{$P_2^*$}] (B) at (1.15,-0.5);
\coordinate [label=below:{$P_3^*$}] (B) at (1.55,-0.5);
\coordinate [label=below:{$P_4^*$}] (B) at (1.9,-0.5);
\coordinate [label=right:{$A$}] (B) at (3.9,3.15);
\coordinate [label=below:{$B$}] (B) at (4.3,1.1);
\coordinate [label=below:{$C$}] (B) at (4.25,2.17);
\coordinate [label=below:{$ $}] (B) at (0.9,0);
\coordinate [label=below:{$ $}] (B) at (1.2,0);
\coordinate [label=below:{$ $}] (B) at (1.5,0);
\coordinate [label=below:{$P_0^{\phantom{*}}$}] (O) at (0,0);
\draw plot[only marks, mark=diamond*, mark size=2pt, mark options={fill=black!30,draw=black!30}] coordinates {(0.75,0)} ;
\draw plot[only marks, mark=diamond*, mark size=2pt] coordinates {(1.15,0) (1.55,0) (1.9,0)} ;
\end{tikzpicture}
\caption{Cyclic stress-softening of a rubber vulcanizate.}
\label{fig:creepcycle2} 
\end{figure}

The time dependency of a cyclically stretched rubber specimen up to a particular strain is represented in Figure \ref{fig:creepcycle2}. The process starts from an unstressed virgin state at $P_0^{\phantom{*}}$ and the stress-strain relation follows path $A$, the primary loading path, until point $P_1$ is reached at a time $t_1^{\phantom{*}}$. At this point $P_1^{\phantom{*}}$, unloading of the rubber specimen begins immediately and the stress-strain relation of the specimen follows the new path $B$ returning to the unstressed state at point $P^*_1$ and time $t^*_1$.  As a result of residual strain, point  $P^*_1$ may not coincide with the origin $P_0^{\phantom{*}}$, but rather be at a position to the right of $P_0^{\phantom{*}}$,   marked by the grey diamond in Figure \ref{fig:creepcycle2}. We assume that reloading  commences immediately, before the onset of recovery or creep of residual strain, and that the stress-strain behaviour subsequently  follows the grey path $C$ until the same maximum strain is reached, at point $P_2^{\phantom{*}}$ and time $t_2^{\phantom{*}}$.  This pattern  then continues throughout the unloading and reloading process as shown in Figure \ref{fig:creepcycle2}. Eventually, an equilibrium state is reached, where the unloading and reloading paths coincide with the previous cycle.
In this paper we do not model creep of residual strain as this appears to play only a small role in the application we discuss in Section \ref{sec:experimental}.  This effect was modelled by the authors in \cite{rickaby2} in the context of biological materials.

We derive here an orthotropic model to represent the Mullins effect for cyclic stress-softening under pure shear deformation.  In pure shear there are three different principal stretches so that the strain-induced  anisotropy of the stress response  is different in each of these three directions, leading to the need for an orthotropic model.   In Section \ref{sec:multi} we describe stress-softening to multiple stress-strain values as initially presented by Rickaby and Scott \cite{rickaby2}. In Section \ref{sec:elasticity} we present a few preliminary definitions on isotropic elasticity.  The orthotropic model is developed in Section \ref{sec:anisotropic} through to Section \ref{sec:constitutive}. Section \ref{sec:anisotropic} follows the work of Spencer \cite{spencer} and provides the foundations of an orthotropic model, which is then continued through Sections \ref{sec:orthotropiceight}, \ref{sec:orthotropic} and \ref{sec:orthresidual} where orthotropic models are derived for the Arruda-Boyce \cite{arruda} eight-chain model, stress-softening and residual strain functions. In Section \ref{sec:softening} we state constitutive models for the softening function and stress softening on the primary loading paths. In Sections \ref{sec:constitutive} and \ref{sec:experimental} we present a constitutive orthotropic model and compare it with experimental data. Finally, in Section \ref{sec:conclusion} we draw some conclusions.

\section{Multiple stress-strain cycles} 
\label{sec:multi}
\begin{figure}[H] 
\centering
\begin{tikzpicture}[scale=1.15, decoration={
markings,
mark=at position 3.2cm with {\arrow[black]{triangle 45};},
mark=at position 11cm with {\arrowreversed[black]{triangle 45};},
mark=at position 21.5cm with {\arrow[black]{triangle 45};},
mark=at position 30cm with {\arrow[black]{triangle 45};},
}
]
\coordinate [label=left:{Stretch, $\lambda$}] (B) at (9,-0.5);
\draw[->] (-0.5,0) -- (11,0)node[sloped,below,midway] {} ;
\draw[->] (-0.5,0) -- (-0.5,7) node[sloped,above,midway] {Stress, $T_{11}$};
\draw[black!80, dotted]
(10,7)--(10,0)
(6.5,7)--(6.5,0)
(3,7)--(3,0);
\draw [black]
(-0.5,0) to[out=50,in=200] node [sloped,above] {} (1.6,1.4)
(1.6,1.4) to[out=20,in=220] node [sloped,above] {} (3.0,2.1)
(0.0,0) to[out=6,in=260] node [sloped,above] {} (3.0,2.1);
\draw [black!40]
(0.0,0) to[out=40,in=240] node [sloped,above] {} (3.0,1.7);
\draw [postaction={decorate}][loosely dotted,line width=0.01pt,black!40]
(-0.5,0) to[out=50,in=200] node [sloped,above] {} (1.6,1.4)
(1.6,1.4) to[out=20,in=220] node [sloped,above] {} (3.0,2.1)
(0,0)--(0,3.6)
(0.0,0) to[out=6,in=260] node [sloped,above] {} (3.0,2.1)
(0,0)--(0,5.4)
(0.0,0) to[out=40,in=240] node [sloped,above] {} (3.0,1.7)
(3.0,2.1) to[out=38,in=215] node [sloped,above] {} (10,6.8);
\fill (3.0,2.1) circle (1.2pt);
\fill (3.0,1.7) circle (1.2pt);
\coordinate [label=above:{$P_1^{\phantom{*}},t_1^{\phantom{*}}$}] (B) at (2.62,2.07);
\coordinate [label=below:{$P_2^{\phantom{*}},t_2^{\phantom{*}}$}] (B) at (3.35,1.74);
\coordinate [label=below:{$t_1^*$}] (B) at (0.0,-0.1);
\coordinate [label=below:{$P_1^*$}] (B) at (0.0,-0.6);
\draw [black]
(3.0,1.7) to[out=53,in=220] node [sloped,above] {} (6.5,4.45)
(2.0,0) to[out=6,in=260] node [sloped,above] {} (6.5,4.45);
\draw [black!40]
(2,0) to[out=45,in=240] node [sloped,above] {} (6.5,3.6);
\fill (6.5,4.45) circle (1.2pt);
\fill (6.5,3.6) circle (1.2pt);
\coordinate [label=above:{$P_3^{\phantom{*}},t_3^{\phantom{*}}$}] (B) at (6.05,4.42);
\coordinate [label=below:{$P_4^{\phantom{*}},t_4^{\phantom{*}}$}] (B) at (6.9,3.6);
\coordinate [label=below:{$t_3^*$}] (B) at (2.0,-0.1);
\coordinate [label=below:{$P_3^*$}] (B) at (2.0,-0.6);
\draw [postaction={decorate}][loosely dotted,line width=0.01pt,black!40]
(3.0,1.7) to[out=53,in=220] node [sloped,above] {} (6.5,4.45)
(0,0)--(0,2.3)
(2.0,0) to[out=6,in=260] node [sloped,above] {} (6.5,4.45)
(0,0)--(0,3.0)
(2.0,0) to[out=45,in=240] node [sloped,above] {} (6.5,3.6);
\draw [black]
(6.5,3.6) to[out=60,in=220] node [sloped,above] {} (10,6.6)
(4.0,0) to[out=6,in=260] node [sloped,above] {} (10,6.6);
\draw [black!40]
(4,0) to[out=50,in=240] node [sloped,above] {} (10,5.5)
(4.5,0) to[out=6,in=260] node [sloped,above] {} (10,5.5);
\fill (10,6.6) circle (1.2pt);
\fill (10,5.5) circle (1.2pt);
\coordinate [label=above:{$P_5^{\phantom{*}},t_5^{\phantom{*}}$}] (B) at (9.5,6.62);
\coordinate [label=below:{$P_6^{\phantom{*}},t_6^{\phantom{*}}$}] (B) at (10.4,5.5);
\coordinate [label=below:{$t_5^*$}] (B) at (4.0,-0.1);
\coordinate [label=below:{$P_5^*$}] (B) at (4.0,-0.6);
\coordinate [label=below:{$t_6^*$}] (B) at (4.5,-0.1);
\coordinate [label=below:{$P_6^*$}] (B) at (4.5,-0.6);
\draw [postaction={decorate}][loosely dotted,line width=0.01pt,black!40]
(6.5,3.6) to[out=60,in=220] node [sloped,above] {} (10,6.6)
(0,0)--(0,0.5)
(4.0,0) to[out=6,in=260] node [sloped,above] {} (10,6.6)
(4,0) to[out=50,in=240] node [sloped,above] {} (10,5.5);
\draw plot[only marks, mark=diamond*, mark size=1.5pt] coordinates {(0.0,0) (2.0,0) (4.5,0) (4,0)} ;
\draw [dashed]
(3.0,2.1) to[out=38,in=215] node [sloped,above] {} (10,6.8);
\coordinate [label=above:{$A$}] (B) at (1.45,1.5);
\coordinate [label=above:{$A'$}] (B) at (4.9,2.75);
\coordinate [label=above:{$A''$}] (B) at (8.3,4.9);
\coordinate [label=above:{$\bar{A}$}] (B) at (7.3,5.2);
\coordinate [label=below:{$P_0^{\phantom{*}}$}] (O) at (-0.5,0);
\end{tikzpicture}
\caption{Cyclic stress-softening to multiple stress-strain cycles.}
\label{fig:tcreepcycle}
\end{figure}
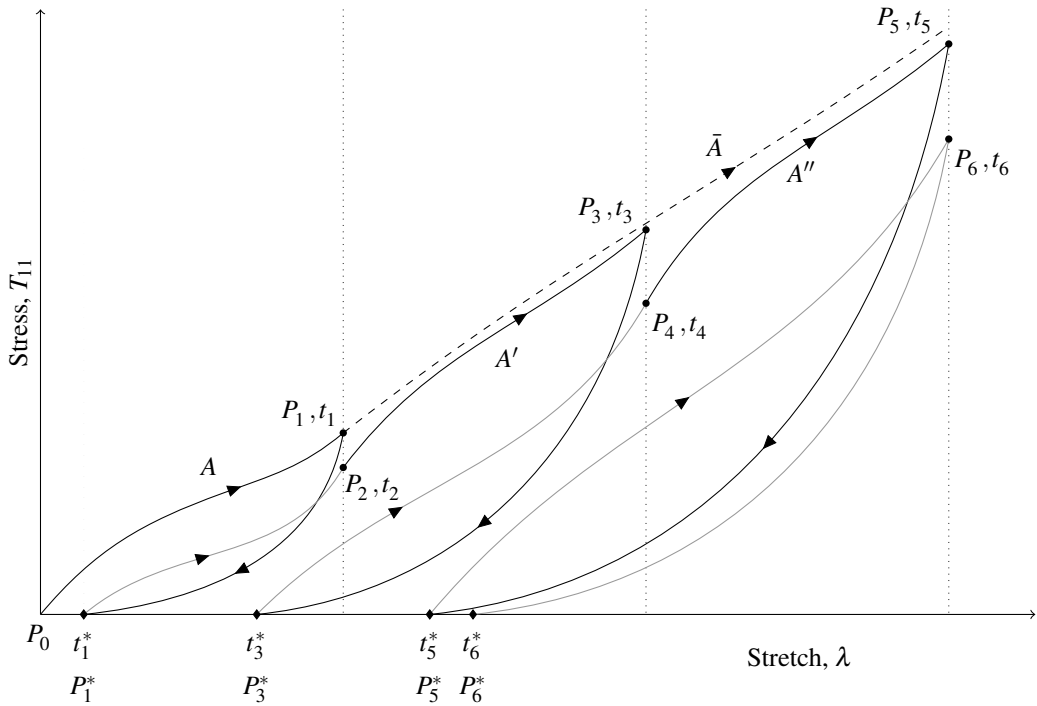

The time-dependent response of a cyclically stretched rubber vulcanizate to multiple strain cycles is represented in Figure \ref{fig:tcreepcycle}. The specimen is loaded along path $A$ to the particular stretch value of $\lambda=\lambda_{\C\_1}$  at point $P_1^{\phantom{*}}$ and corresponding time $t_1^{\phantom{*}}$.   This is the commencement of cycle one and $\lambda_{\C\_1}$ is the maximum stretch value for cycle one.  Unloading of the rubber specimen begins immediately  from point $P_1^{\phantom{*}}$ and the material returns to the unstressed state at point $P^*_1$ and time $t^*_1$.  
Reloading then commences immediately, ceasing when the same stretch value $\lambda=\lambda_{\C\_1}$ is reached once more, this time at the different point  $P_2^{\phantom{*}}$ and time   $t_2^{\phantom{*}}$. 
The material is immediately stretched beyond the strain value $\lambda=\lambda_{\C\_1}$ along the first new primary loading path $A'$ to a new maximum stretch $\lambda=\lambda_{\C\_2}$  at the point $P_3^{\phantom{*}}$ and time $t_3^{\phantom{*}}$.   This is the start of cycle two and $\lambda_{\C\_2}$ is the maximum stretch value for this cycle. The specimen is then unloaded to zero stress at the point $P^*_3$ and time $t^*_3$.    
Reloading then commences immediately, ceasing when the same stretch value $\lambda=\lambda_{\C\_2}$ is reached once more, this time at the different point  $P_4^{\phantom{*}}$ and time   $t_4^{\phantom{*}}$. 
The material is immediately stretched beyond the strain value $\lambda=\lambda_{\C\_2}$ along the second new primary loading path $A''$ to a new maximum stretch $\lambda=\lambda_{\C\_3}$  at the point $P_5^{\phantom{*}}$ and time $t_5^{\phantom{*}}$.   This is the start of cycle three and $\lambda_{\C\_3}$ is the maximum stretch value for this cycle. The specimen is then unloaded to zero stress at the point $P^*_5$ and time $t^*_5$.    It is then reloaded to the same stretch value  $\lambda=\lambda_{\C\_3}$ at point  $P_6^{\phantom{*}}$ and time   $t_6^{\phantom{*}}$ and so the process goes on. 
These observations are borne out from the experimental data of Diani \textit{et al.} \cite[Figure 1]{dianib}.  For further details on the concept of multiple  stress-strain cyclic loading, see Rickaby and Scott \cite[Section 2]{rickaby2}.

\section{Preliminary functions} 
\label{sec:elasticity}

In the reference configuration, at time $t_0^{\phantom{*}}$, a material particle is  located at $\textbf{X}$ with Cartesian components $X_1,X_2,X_3$, relative to the orthonormal Cartesian basis $\{{\bf e}_1, {\bf e}_2, {\bf e}_3\}$.   After  deformation, at time $t$, the same particle is located  at the position $\bm{x}(\textbf{X},t)$ with components $x_1,x_2,x_3$, 
relative to the same orthonormal basis $\{{\bf e}_1, {\bf e}_2, {\bf e}_3\}$.   The deformation gradient is defined by
\begin{equation*}
F_{iA}(\textbf{X},t)=\frac{\partial x_i(\textbf{X},t)}{\partial X_A}.
\end{equation*}
A pure shear strain is taken in the form
\begin{equation}
x_1=\lambda X_1, \quad x_2=\lambda^{-1} X_2, \quad x_3=X_3,
\label{eq:1yy}
\end{equation}
where $\lambda>1$ is the greatest principal stretch.

The left and right Cauchy-Green strain tensors $\textbf{B}=\textbf{F}\textbf{F}^\T$ and 
$\textbf{C}=\textbf{F}^\T\textbf{F}$, respectively, are given by
\begin{displaymath}
\textbf{B}= \left( \begin{array}{ccc}	
{\lambda^2}& 0 &0\\
0 &{\lambda^{-2}}& 0\\
0& 0 &1 \end{array}\right), \quad
\textbf{C}= \left( \begin{array}{ccc}	
{\lambda^2}& 0 &0\\
0 &{\lambda^{-2}}& 0\\
0 & 0 & 1 \end{array}\right),
\end{displaymath}
and are equal.  They have common
 principal invariants
\begin{equation}
 I_1 = \tr {\bf C} = \lambda^2+\lambda^{-2}+1,\qquad I_2 = I_3 \,\tr {\bf C}^{-1} = \lambda^2+\lambda^{-2}+1,
\qquad I_3 =\det {\bf C}= 1,
\label{eq:2yy}
\end{equation}
the last being a consequence of isochoricity.

An incompressible isotropic hyperelastic material possesses a strain energy function $W(I_1, I_2)$ in terms of which the Cauchy stress is given by
\begin{align}
\textbf{T}^{\mathscr{E}_{\iso}}(\lambda) =& \, -p\textbf{I}+2\left[\frac{\partial{W}}{\partial I_1}+I_1\frac{\partial{W}}{\partial I_2}\right]
\textbf{B}-2\frac{\partial{W}}{\partial I_2}\textbf{B}^2,
\label{eq:3yy}
\end{align}
where the superscript $\mathscr{E}_{\iso}$ refers to isotropic elasticity and $\bf I$ is the identity tensor.   The arbitrary pressure  $p$ is  fixed by the requirement 
 $\textbf{T}^{\mathscr{E}_{\iso}}_{22}(\lambda)=0$ to be
\[
p  =  2\frac{\partial{W}}{\partial I_1}\lambda^{-2} + 2\frac{\partial{W}}{\partial I_2}(1+\lambda^{-2}).
\]
Equation (\ref{eq:3yy}) then gives the two non-zero stress components in  pure shear to be
\begin{align}
 T_{11}^{\mathscr{E}_{\iso}}(\lambda) &=   2(\lambda^2- \lambda^{-2})
 \left[\frac{\partial{W}}{\partial I_1}  +  \frac{\partial{W}}{\partial I_2}\right],
\label{eq:4yy} \\
 T_{33}^{\mathscr{E}_{\iso}}(\lambda)  &=    2 (\lambda^{2}-1)
 \left[\lambda^{-2}\frac{\partial{W}}{\partial I_1} +  \frac{\partial{W}}{\partial I_2}\right].
\label{eq:5yy}
\end{align}
Assuming that the empirical inequalities
\[ \frac{\partial{W}}{\partial I_1}>0,\qquad \frac{\partial{W}}{\partial I_2}\geq 0\] 
hold, we see that
 $T_{11}^{\mathscr{E}_{\iso}}>0$ and $T_{33}^{\mathscr{E}_{\iso}}>0$ because  $\lambda>1$. Additional details on isotropic stress-softening in pure shear may be found in Beatty \cite{beattyb}.

The Arruda-Boyce \cite{arruda} isotropic eight-chain model has strain energy
\begin{equation}
W_{\iso}=\mu N \left\{\beta\mathscr{L}(\beta) +\log \left(\frac{{\beta}}{\sinh{\beta}}\right) \right\},
\label{eq:6yy}
\end{equation}
where
\begin{equation*}
{\beta}=\mathscr{L}^{-1}\left(\frac{{{\lambda}_{\chain}}}{\sqrt N}\right) \quad \with \quad \lambda_{\chain}=\sqrt{\frac{I_1}{{3}}},
\end{equation*}
and $\mu$ is a shear modulus.   $N$ is the number of links forming a single polymer chain and $y=\mathscr{L}^{-1}(x)$ is the inverse Langevin function where the Langevin function is defined by
\[
x=\mathscr{L}(y)=\coth y-\frac{1}{y}.
\]
Upon substituting for $W$ from equation (\ref{eq:6yy}) into equation (\ref{eq:3yy}) we obtain the stress in the  Arruda-Boyce model of isotropic elasticity:
\begin{align}
\textbf{T}^{\mathscr{E}_{\iso}}(\lambda) =& \, -p \textbf{I}+ \mu \sqrt{\frac{N}{3I_1}}\mathscr{L}^{-1}\left(\sqrt{\frac{I_1}{3{N}}}\right)\textbf{B}.
\label{eq:7yy}
\end{align}

A standard, simple approximation to the inverse Langevin function, often used in the literature, is that of Cohen \cite{cohen}:
\begin{equation}
\mathscr{L}^{-1}(x)\approx 3x\frac{1- \tfrac13x^2}{1-x^2},
\label{eq:cohen}
\end{equation}
valid for $|x|<1$, which is an approximation to a certain Pad\'e approximant of  $\mathscr{L}^{-1}(x)$.
For uniaxial strain, the good agreement between the isotropic elastic stress calculated using the inverse Langevin function and that using Cohen's approximation (\ref{eq:cohen}) is noted, for example, by Rickaby and Scott \cite{rickaby4} in the context of uniaxial compression.

 Rickaby and Scott \cite{rickaby5} propose the new approximation
\begin{equation}
\mathscr{L}^{-1}(x)\approx 3x\frac{1- \tfrac25x^2}{1-x^2},
\label{eq:new}
\end{equation}
which is as simple as Cohen's but a more accurate approximation to  $\mathscr{L}^{-1}(x)$ over most of the $x$ range. 
For example, the mean percentage error over the range $0<x<0.95$ of Cohen's approximation (\ref{eq:cohen}) is $2.74\%$, whilst that of (\ref{eq:new}) is only $0.32\%$.  Therefore, when comparing the model to experimental data  in Section \ref{sec:experimental} of this paper, we employ the approximation (\ref{eq:new}) for $\mathscr{L}^{-1}(x)$.

\section{Orthotropic elastic response} 
\label{sec:anisotropic}
For the pure shear deformation (\ref{eq:1yy}),  a tension (\ref{eq:4yy}) is applied in the 1-direction, so that $\lambda>1$, and a compression (\ref{eq:4yy}) is applied in the 2-direction.  This generates two preferred material directions,  the 1,2-directions of the extension and compression, respectively. These preferred directions are recorded by the material and influence the subsequent response of the material. If loading is terminated at a certain strain $\lambda_{\C\_1}$, then the damage caused is now dependent on the value of strain $\lambda_{\C\_1}$;  this must be reflected in the response of the material upon unloading and subsequent submaximal reloading. The material response must now therefore be regarded as  orthotropic relative to the original reference configuration. 

Spencer \cite{spencer} characterized an orthotropic elastic solid by the existence of two preferred directions, denoted by the unit vector fields $\textbf{u}(\textbf{X})$ and $\textbf{v}(\textbf{X})$.  After deformation the preferred directions 
$\textbf{u}(\textbf{X})$ and $\textbf{v}(\textbf{X})$ become parallel to
 \[\bm{a}=\mathbf{Fu}, \quad \bm{b}=\mathbf{Fv}, \]
which are not in general unit vectors.

The strain energy $W$ is now described by $W(I_{1}, \ldots,I_{10})$, with the invariants $I_1$ to $I_3$ being defined by (\ref{eq:2yy}) and $I_4$ to $I_{10}$ being given by,
\begin{align}
& I_4=\textbf{u}\cdot(\textbf{C}\textbf{u}),\quad 
I_5=\textbf{u}\cdot(\textbf{C}^2\textbf{u}),\quad
I_6=\textbf{v}\cdot(\textbf{C}\textbf{v}),\quad
I_7=\textbf{v}\cdot(\textbf{C}^2\textbf{v}),  \nonumber
\\
& I_8=(\textbf{u}\cdot\textbf{v})\textbf{u}\cdot(\textbf{C}\textbf{v}),\quad 
I_9=(\textbf{u}\cdot\textbf{v})\textbf{u}\cdot(\textbf{C}^2\textbf{v}),\quad
I_{10}=(\textbf{u}\cdot\textbf{v})^2.
\label{eq:8yy}
\end{align}
An identity relating these ten invariants may be written 
\begin{align}
&\frac{1}{2}(\textbf{u} \times \textbf{v})\cdot(\textbf{u} \times \textbf{v})\left\{(\tr\textbf{C})^2-\tr\textbf{C}^2\right\}+2(\textbf{u}\cdot\textbf{v})\left\{(\textbf{u}\cdot(\textbf{C}\textbf{v}))\tr\textbf{C}-\textbf{u}\cdot(\textbf{C}^2\textbf{v})\right\}-(\textbf{u}\cdot(\textbf{C}\textbf{v}))^2   \nonumber\\
&\;\;{} -\, \left\{\textbf{u}\cdot(\textbf{C}\textbf{u})
+\textbf{v}\cdot(\textbf{C}\textbf{v})\right\}\tr\textbf{C}+(\textbf{u}\cdot(\textbf{C}\textbf{u}))(\textbf{v}\cdot(\textbf{C}\textbf{v}))+\textbf{u}\cdot(\textbf{C}^2\textbf{u})+\textbf{v}\cdot(\textbf{C}^2\textbf{v})=0,
\label{eq:9yy}
\end{align}
the derivation of which is provided in Appendix A.   Spencer \cite[eqn (33)]{spencer} presents this identity but omits the factor of $1/2$ in the leading term.   This identity may also be written purely in terms of $I_1,\dots,I_{10}$ as
\begin{equation}
(1-I_{10})I_2+2I_8I_1-2I_9-I_{10}^{-1}I_8^2-I_4I_1-I_6I_1+I_4I_6+I_5+I_7 =0.
\label{eq:9yyy}
\end{equation}
From the identity (\ref{eq:9yyy}) it is clear that we may omit, say, the invariant $I_9$ from the list of arguments of the strain energy function $W$.  We may also omit $I_{10}$ as this does not give rise to a stress.  The elastic stress in an incompressible orthotropic elastic material is then given in terms of $W(I_1,\dots,I_8)$  by
\begin{align}
\textbf{T}^{\mathscr{E}_{\oi}} =-p\textbf{I}+2\bigg\{&\, \left(\frac{\partial{W}}{\partial I_1}+I_1
\frac{\partial{W}}{\partial I_2}\right)
\textbf{B}-\frac{\partial{W}}{\partial I_2}\textbf{B}^2\nonumber\\
&\mbox{}+ \frac{\partial{W}}{\partial
I_4}\bm{a}\otimes\bm{a}+\frac{\partial{W}}{\partial
I_5}[\bm{a}\otimes\textbf{B}\bm{a}+\textbf{B}\bm{a}\otimes\bm{a}]+ \frac{\partial{W}}{\partial
I_6}\bm{b}\otimes\bm{b}\nonumber\\
&+\frac{\partial{W}}{\partial
I_7}[\bm{b}\otimes\textbf{B}\bm{b}+\textbf{B}\bm{b}\otimes\bm{b}]\bigg\}+\frac{\partial{W}}{\partial
I_8}[\bm{a}\otimes\bm{b}+\bm{b}\otimes\bm{a}],
\label{eq:10yy}
\end{align}
where $\otimes$ denotes a dyadic product and the superscript $\mathscr{E}_{\oi}$ refers to orthotropic elasticity.  

The preferred direction $\textbf{u}$ lies in the 1-direction of the deformation (\ref{eq:1yy}), so that
\begin{equation}
\textbf{u} = {\bf e}_1 = \left(\begin{array}{c}1\\0\\0\end{array}\right),\quad
\bm{a} =  \left(\begin{array}{c}\lambda \\0\\0\end{array}\right),\quad
\bm{a}\otimes \bm{a}
= \left( \begin{array}{ccc}	
\lambda^2& 0 & 0\\
0 & 0 & 0\\
0 & 0 & 0 \end{array}\right)
,\quad \bm{Ba}=\lambda^2\bm{a}.
\label{eq:11yy}
\end{equation}
The preferred direction $\textbf{v}$ lies in the 2-direction of the deformation (\ref{eq:1yy}), so that,
\begin{equation}
\textbf{v} ={\bf e}_2 = \left(\begin{array}{c}0\\1\\0\end{array}\right),\quad
\bm{b} = \left(\begin{array}{c} 0\\ \lambda^{-1} \\0\end{array}\right),\quad
\bm{b}\otimes \bm{b}
= \left( \begin{array}{ccc}	
0 & 0 & 0\\
0 & \lambda^{-2} & 0\\
0 & 0 & 0 \end{array}\right)
, \quad \bm{Bb}=\lambda^{-2}\bm{b}.
\label{eq:12yy}
\end{equation}
We have taken the preferred directions $\textbf{u}$ and $\textbf{v}$ of orthotropicity to be perpendicular, so that 
$\textbf{u}\cdot\textbf{v}=0$, and so $I_8=I_9=I_{10}=0$ and from equations (\ref{eq:11yy}) and (\ref{eq:12yy})  the remaining anisotropic invariants are
\begin{equation}
I_4=\lambda^2,\quad I_5=\lambda^4,\quad I_6=\lambda^{-2},\quad I_7=\lambda^{-4}. 
\label{eq:4.7a}\end{equation}
For this choice of invariants, we can see that identity (4.3) is satisfied. This is consistent with the work of other authors, including Spencer \cite{spencer}, Holzapfel \cite[pages 274-275 ]{holzapfel} and Ogden \cite[pages 192-193]{ogdenbook}.

We shall see in the next section that in the orthotropic  Arruda-Boyce model only the invariants  $I_1, I_4, I_6$ are involved 
and so
our final form of the strain energy is therefore $W=W(I_1,I_4,I_6)$, giving rise from  (\ref{eq:10yy}) to the stress
\begin{equation}
\textbf{T}^{\mathscr{E}_\oi} = -p\textbf{I}+2\bigg\{\,\frac{\partial{W}}{\partial I_1}\textbf{B}
+ \frac{\partial{W}}{\partial I_4}\bm{a}\otimes\bm{a} + \frac{\partial{W}}{\partial I_6}\bm{b}\otimes\bm{b}\bigg\},
\label{eq:13yy}
\end{equation}
which is equivalent to the constitutive equation of Spencer \cite[eqn (71)]{spencer} for an incompressible orthotropic elastic material with the invariant $I_2$ removed.

\section{Orthotropic eight-chain model of elasticity} 
\label{sec:orthotropiceight}

We extend the work of Kuhl \textit{et al.} \cite{kuhl} and Bischoff \textit{et al.} \cite{Bischoff}  in order to develop a simple model for orthotropic elasticity based on the original  Arruda-Boyce \cite{arruda} eight-chain  model of  isotropic elasticity. Rubber is regarded  as being composed of cross-linked polymer chains, each chain consisting of $N$ links, with  each link being of length $l$. The two parameters, $N$ and $l$  are related through the locking length $ r_{\rm L}^{\phantom{L}}$ and chain vector length $r_0$, where
\begin{equation}
 r_{\rm L}^{\phantom{L}} = Nl,\qquad r_0 = \sqrt{N}l.
\label{eq:16yy}
\end{equation}
The locking length $ r_{\rm L}^{\phantom{L}}$ is the length of the polymer chain when fully extended.  
The chain vector length $r_0$ is the distance between the two ends of the chain in the undeformed configuration. 
Due to significant coiling of the polymer chains this length is considerably less than the locking length.
The value $ r_0 = \sqrt{N}l$ is derived by statistical considerations.

\begin{figure}[h] 
\centering
\begin{tikzpicture}[scale=1.00]
%
\draw [black!40,decorate,decoration=snake] 
(3,6) to [out=-120,in=120] node [sloped,below]{} (4.5,3.5);
\draw [black!40,decorate,decoration=snake]
(4.5,3.5) to [out=100,in=-200] node [sloped,below]{} (5,4)
(5,1) to [out=140,in=-60] node [sloped,below]{} (4.5,3.5)
(1,1) to [out=30,in=210] node [sloped,below]{} (4.5,3.5)
(3,3) to [out=65,in=285] node [sloped,below]{} (4.5,3.5);
\draw [black!40,decorate,decoration=snake]
(7,3) to [out=140,in=340] node [sloped,below]{} (4.5,3.5)
(4.5,3.5) to [out=140,in=0] node [sloped,below]{} (1,4)
(4.5,3.5) to [out=10,in=230] node [sloped,below]{} (7,6);
\draw [->](1,1) -- (0,0);
\draw [->](7,3) to[out=0,in=180] node [sloped,above] {} (8,3);
\draw [->](3,6) to[out=90,in=270] node [sloped,above] {} (3,7);
\draw [postaction={decorate}]
(0,0) to [out=45,in=225] node [sloped,below]{} (1,1)
(5,1) to [out=45,in=225] node [sloped,below]{} (7,3)
(1,4) to [out=45,in=225] node [sloped,below]{} (3,6)
(5,4) to [out=45,in=225] node [sloped,below]{} (7,6)
(7,3) to[out=0,in=180] node [sloped,above] {} (8,3)
(1,4) to[out=0,in=180] node [sloped,above] {} (5,4)
(1,1) to[out=0,in=180] node [sloped,above] {} (5,1)
(3,6) to[out=0,in=180] node [sloped,above] {} (7,6)
(1,1) to[out=90,in=270] node [sloped,above] {} (1,4)
(5,1) to[out=90,in=270] node [sloped,above] {} (5,4)
(3,6) to[out=90,in=270] node [sloped,above] {} (3,7)
(7,3) to[out=90,in=270] node [sloped,above] {} (7,6);
\coordinate [label=above:{$x_3$}] (A) at (0.4,-0.2);
\coordinate [label=left:{$x_1$}] (B) at (8.2,2.7);
\coordinate [label=below:{$x_2$}] (C) at (3.3,7.2);
\coordinate [label=above:{$a$}] (D) at (3,0.2);
\coordinate [label=left:{$c$}] (E) at (6.8,2);
\coordinate [label=below:{$b$}] (F) at (7.5,4.7);
%
\draw[dashed][postaction={decorate}]
(1,1) to [out=45,in=225] node [sloped,below]{} (3,3)
(3,3) to[out=0,in=180] node [sloped,above] {} (7,3)
(3,3) to[out=90,in=270] node [sloped,above] {} (3,6);
\fill [black!40] (7,6) circle (4pt);
\fill [black!40] (3,6) circle (4pt);
\fill [black!40] (3,3) circle (4pt);
\fill [black!40] (7,3)circle (4pt);
\fill [black!40] (1,1)circle (4pt);
\fill [black!40] (5,1)circle (4pt);
\fill [black!40] (5,4)circle (4pt);
\fill [black!40] (1,4)circle (4pt);
\fill [black!40] (4.5,3.5)circle (4pt);
\draw [stealth-stealth,line width=1.3pt](4.5,3.5) -- (1,4);
\draw [stealth-stealth,line width=1.3pt](1,0.7) -- (5,0.7);
\draw [stealth-stealth,line width=1.3pt](5.3,1) -- (7.3,3);
\draw [stealth-stealth,line width=1.3pt](7.3,3) -- (7.3,6);
\coordinate [label=below:{$r_0$}] (A) at (2.7,3.6);
%
%
%
\end{tikzpicture}
\caption{The orthotropic Arruda-Boyce  eight-chain model.  The cube of the isotropic case is replaced by a cuboid with generally unequal sides $a$, $b$, $c$.} 
\label{fig:arruda4}  
\end{figure}
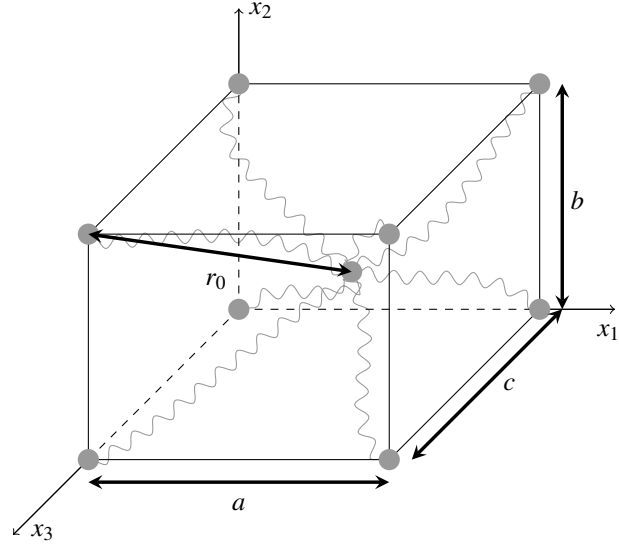

In this extension of the Arruda-Boyce model we consider a cuboid aligned with its edges parallel to the coordinate axes, as in Figure \ref{fig:arruda4}.  The edges parallel to the $x_1$, $x_2$-axis, are considered to be the preferred orthotropic material directions, with lengths $a$ and $b$, respectively. The remaining edge is then of length $c$.  Each of the eight vertices of the cuboid is attached to the centre point of the cuboid by a polymer chain, as depicted in Figure \ref{fig:arruda4}.  Each of these eight chains is of the same length in the undeformed state which we take to be the vector chain length $r_0$.

From  Figure \ref{fig:arruda4}  we see that   the chain vector length may be written
\begin{equation}
r_0=\sqrt{\left(\frac{1}{2}a\right)^2+\left(\frac{1}{2}b\right)^2+\left(\frac{1}{2}c\right)^2}.
\label{eq:17yy}
\end{equation}
We consider a triaxial stretch along the coordinate axes $\{ {\bf e}_1,  {\bf e}_2,  {\bf e}_3 \}$ with principal stretches, 
$\lambda_1, \lambda_2, \lambda_3$, respectively.  The cuboid is not rotated by this deformation but now has sides of lengths $a\lambda_1, b\lambda_2, c\lambda_3$, respectively.   Thus, the deformed length of each of the eight chains is  given by
\begin{equation*}
r_\chain=\sqrt{\left(\frac{1}{2}a\lambda_1\right)^2+\left(\frac{1}{2}b\lambda_2\right)^2+\left(\frac{1}{2}c\lambda_3\right)^2}.
\end{equation*}
Taking ${\bf u}={\bf e}_1$ and ${\bf v}={\bf e}_2$ we see from (\ref{eq:2yy})$_{1}$ and  (\ref{eq:8yy})$_{1,3}$ that
\[ I_1= \lambda_1^2 +  \lambda_2^2 +  \lambda_3^2,\quad I_4 = \lambda_1^2,\quad  I_6= \lambda_2^2, \]
from which it follows that 
\[  \lambda_1^2 = I_4 ,\quad  \lambda_2^2 =  I_6,\quad  \lambda_3^2 = I_1-I_4-I_6. \]
Therefore,  $r_\chain$ may be written 
\begin{equation}
r_\chain=\frac{1}{2}\sqrt{I_4a^2+I_6b^2 +\left[I_1-I_4-I_6\right]c^2}.
\label{eq:18yy}
\end{equation}
The argument of the inverse Langevin function is given by
\[\frac{r_\chain}{r_L}\]
where $r_L$ is given in equation (\ref{eq:16yy})$_1$.  We have, using equations (\ref{eq:16yy})$_2$, (\ref{eq:17yy}) and (\ref{eq:18yy}),
\begin{align*}
\frac{r_\chain}{r_L} = \frac{r_\chain}{r_0}\cdot\frac{r_0}{Nl} = \,&
\frac{\sqrt{I_4a^2+I_6b^2 +\left[I_1-I_4-I_6\right]c^2}}{\sqrt{a^2+b^2+c^2}}\cdot \frac{\sqrt{N}l}{Nl} \nonumber\\
= \,& \sqrt{\frac{I_4a^2+I_6b^2 +\left[I_1-I_4-I_6\right]c^2}{N(a^2+b^2+c^2)}}.  
\end{align*}
The quantity $\beta$ is defined by
\begin{align}
\beta
=\mathscr{L}^{-1}\left(\frac{r_\chain}{r_L}\right) = \,&
\mathscr{L}^{-1}\left(\sqrt{\frac{I_4a^2+I_6b^2 +\left[I_1-I_4-I_6\right]c^2}{N(a^2+b^2+c^2)}} \right)\nonumber\\ 
= \,&
\mathscr{L}^{-1}\left(\sqrt{\frac{I_4+I_6\alpha^2_1 +\left[I_1-I_4-I_6\right]\alpha^2_2}{N(1+\alpha^2_1+\alpha^2_2)}} 
\right) \nonumber\\
 = \,& \mathscr{L}^{-1}(\gamma), \label{eq:19yy} \\
 \intertext{where the argument of the inverse Langevin function $\gamma$ is defined by}
  \gamma= \, & \sqrt{\frac{I_4+I_6\alpha^2_1 +\left[I_1-I_4-I_6\right]\alpha^2_2}{N(1+\alpha^2_1+\alpha^2_2)}}.
\label{eq:gamma}
\end{align}
The quantities  $\alpha_1 = b/a$ and $\alpha_2 = c/a$ are the aspect ratios of the cuboid in this extended Arruda-Boyce model. Selecting
$\alpha_1 =\alpha_2 = 1$ in equation (\ref{eq:gamma}) corresponds to material isotropy so that $I_4$, $I_6$ cancel out and we obtain
\begin{equation}
\beta =\mathscr{L}^{-1}\left(\sqrt\frac{I_1}{3 N }\right),
\label{eq:20yy}
\end{equation}
which is consistent with the isotropic Arruda-Boyce \cite{arruda} eight-chain model, see equation (\ref{eq:6yy}).

Substituting equation (\ref{eq:19yy}) into equation (\ref{eq:6yy}) leads to the following orthotropic strain energy:
\begin{align}
W_{\textrm{A-B}}=&\,\mu N \left\{\gamma \mathscr{L}^{-1}(\gamma)+\log \left(\frac{\mathscr{L}^{-1}(\gamma)}{\sinh\left(\mathscr{L}^{-1}(\gamma)\right)}\right) \right\}- \frac{1}{2}h_{4}(I_4-1) - \frac{1}{2}h_{6}(I_6-1),
\label{eq:21yy}
\end{align}
where $h_4$ and $h_6$ are constants chosen so that the  stress vanishes in the  undeformed state. 

Employing the strain energy (\ref{eq:21yy}) in the stress (\ref{eq:13yy}) leads to the following expression for the elastic stress in our orthotropic Arruda-Boyce model:
\[
\textbf{T}^{\mathscr{E}_\oi}(\lambda) =  -p\textbf{I}+2\bigg\{\,\frac{\partial \gamma}{\partial I_1}\frac{\partial {W_{\textrm{A-B}}}}{\partial  \gamma}\textbf{B}
+ \left( \frac{\partial \gamma}{\partial I_4} \frac{\partial {W_{\textrm{A-B}}}}{\partial  \gamma}-\frac12h_4\right)\bm{a}\otimes\bm{a} + \left( \frac{\partial \gamma}{\partial I_6} \frac{\partial {W_{\textrm{A-B}}}}{\partial  \gamma} - \frac12h_6\right)\bm{b}\otimes\bm{b}\bigg\},
 \]
where $\gamma$ is defined by (\ref{eq:gamma}).  This leads to 
\begin{align}
 \mathbf{T}^{\mathscr{E}_{\oi}}(\lambda) = & -p\mathbf{I} + 
\mu\frac{1}{1+\alpha_1^2+\alpha_2^2} \gamma^{-1}\beta \bigg\{\alpha^2_2\mathbf{B} +
(1-\alpha^2_2)\bm{a}\otimes\bm{a}+(\alpha_1^2-\alpha^2_2)\bm{b}\otimes\bm{b}\bigg\}\nonumber\\  
&\quad - h_4\bm{a}\otimes\bm{a}
 - h_6\bm{b}\otimes\bm{b}. 
\label{eq:22yy}
\end{align}
For this stress to vanish in the reference configuration, where  $I_4=I_6=1$ and $\gamma = \sqrt{\frac{1}{N}}$, we must take
\[
h_4 = \mu  \frac{1- \alpha^2_2}{1+\alpha^2_1+\alpha^2_2} \sqrt N  \mathscr{L}^{-1}\left(\sqrt{\frac{1} N }\right),\quad
h_6 = \mu  \frac{\alpha^2_1 - \alpha^2_2}{1+\alpha^2_1+\alpha^2_2} \sqrt N  \mathscr{L}^{-1}\left(\sqrt{\frac{1} N }\right). 
\]
For an isotropic material, $\alpha_1=\alpha_2=1$ and we find that $h_4=h_6=0$,  as expected.

This appears to be the first time that the simple Arruda-Boyce-type  model (\ref{eq:21yy}) for orthotropic elasticity has appeared in the literature. This development follows naturally from the transversely isotropic model  presented by Rickaby and Scott \cite{rickaby1}. In Section \ref{sec:experimental} the model is found to fit the experimental data very well.

For the eight polymer chains to remain equal in length in the Arruda-Boyce-type models of elasticity the edges of the cube or cuboid must be chosen parallel to the  principal axes of the deformation, otherwise the eight chains will not all be the same length after deformation. Therefore the current model is restricted to situations where the principal axes of strain remain fixed throughout the deformation, so that the Arruda-Boyce cube or cuboid may be selected with edges parallel to these principal axes.  The present example of pure shear is a case in point but it is not clear how these methods could be extended, for example, to simple shear.

\section{Softening function} 
\label{sec:softening}

\subsection{Stress softening on the initial primary loading path}
\label{sec:primary}

For carbon-filled vulcanized rubber it is noted that during initial primary loading at very small deformations pronounced softening occurs, see Mullins \cite{mullins1969}. To account for this feature, Rickaby and Scott \cite{rickaby1} introduced the following damage function:  
\begin{equation}
\zeta_{1,0}(\lambda)=\left[1-\frac{1}{r_{\C\_1}}\left\{\tanh\left(\frac{\lambda_{\C\_1}-{\lambda}}{ b_0}\right)\right\}^{{1}/{\vartheta_0}}\right] \quad \for \quad 1\leq\lambda\leq\lambda_{\C\_1}
\label{eq:5.11zzz}
\end{equation} 
where ${r_{\C\_1}}$, $b_0$ and  $\vartheta_0$ are positive constants, with $\lambda_{\C\_1}$ being the greatest stretch achieved on the initial primary loading path. Choosing $\left|r_{\C\_1}\right|\geq 1$ guarantees that $\zeta_{1,0}(\lambda)>0$ for $\lambda \geq 1$ on primary loading.

For initial primary loading, equation (\ref{eq:5.11zzz}) is coupled with the isotropic component of the elastic stress $\textbf{T}^{\mathscr{E}_\iso}(\lambda)$ to give
\[
\textbf{T}^{\mathscr{E}_{\iso}}(\lambda)=\zeta_{1,0}(\lambda)\left[-p\textbf{I}+\left[\mu \sqrt{\frac N {3I_1}}\mathscr{L}^{-1}\left(\sqrt{\frac{I_1}{3{ N }}}\right)\right]\textbf{B}\right].
\]

\subsection{Softening on the unloading and reloading paths}
For softening on the unloading and reloading paths Rickaby and Scott \cite{rickaby3} developed the following softening function:
\begin{equation}
\zeta_{n,\omega}(\lambda) = 1-\frac{1}{r_\omega}\left\{\tanh\left(\frac{{W_{\C\_n}}-{W}}{\mu b_\omega}\right)\right\}^{{1}/{\vartheta_\omega}},
\label{eq:23yy}
\end{equation}
here ${W}$ is the current strain energy value, $W_{\C\_n}$ is the maximum strain energy value achieved on the loading path before unloading with $n$ denoting the cycle number, i.e. in Figure \ref{fig:tcreepcycle} when path $A$ ceases $W_{\C\_n}=W_{\C\_1}$, similarly when path $A'$ ceases $W_{\C\_n}=W_{\C\_2}$. In equation (\ref{eq:23yy}) $b_\omega$, $r_\omega$ are positive dimensionless material constants with $\omega$ being defined by 
\begin{equation}
\omega=
1 \quad \textrm{for unloading},\qquad
\omega = 2 \quad \textrm{for reloading}.
\label{eq:24yy}
\end{equation}

The softening function  (\ref{eq:23yy}) has the property that
 \begin{equation}\label{eq:6.4a}
\textbf{T}^\oi =  \zeta_{n,\omega}(\lambda) \textbf{T}^{\mathscr{E}_\oi}(\lambda),
 \end{equation}
thus providing a relationship between the orthotropic Cauchy stress $\textbf{T}^\oi$ and the orthotropic elastic response, $\textbf{T}^{\mathscr{E}_\oi}(\lambda)$, during unloading and reloading of the material.
The modelling approach of combining the softening function with the stress response, as exemplified by equation (\ref{eq:6.4a}) here, was introduced by Ogden and Roxburgh \cite{ogden} and described by Dorfmann and Ogden \cite{dorfmann2003,dorfmann}, and has subsequently been used by several authors. This modelling approach has been found to significantly improve the accuracy of the fit achieved with experimental data, see Rickaby and Scott \cite{rickaby3,rickaby4}.

\section{Orthotropic stress relaxation} 
\label{sec:orthotropic}
Bernstein \textit{et al.} \cite{bernstein} developed a model  for non-linear stress relaxation which has been found to  represent accurately experimental data for stress-relaxation, see 
Tanner \cite{tanner} and the references therein.

For an orthotropic incompressible viscoelastic solid, we can build on the work of Lockett \cite[pages 114--116] {lockett} and Wineman \cite[Section 12]{Wineman2009}  
to write down the following version of the Bernstein \textit{et al.} \cite{bernstein} model for the relaxation stress
 $\textbf{T}^{\mathscr{R}_{\oi}}$ in an orthotropic material:
\begin{align}
\textbf{T}^{\mathscr{R}_{\oi}}(\lambda,t) & =-p\textbf{I} +\bigg[ {A}_0+\frac{1}{2} \hat{A}_1(t)(I_1-3)-
\hat{A}_2(t)\bigg]{\textbf{B}}+ \hat{A}_2(t){\textbf{B}}^{2}\nonumber\\
&\;\;+\hat{A}_4(t)(I_4-1)\bm{a}\otimes \bm{a}+\hat{A}_6(t)(I_6-1)\bm{b}\otimes \bm{b},
\label{eq:25yy}
\end{align}
for  $ t>t_0^{\phantom{*}}$.  The superscript $\mathscr{R}_{\oi}$ refers to stress relaxation in an orthotropic material.  
As earlier with elasticity theory, we have omitted all anisotropic invariants other than $I_4$ and $I_6$. 
The first line of (\ref{eq:25yy})  is that derived by Lockett \cite[pages 114--116] {lockett} for full isotropy, as given by
\begin{align}
\textbf{T}^{\mathscr{R}_{\iso}}(\lambda,t) & =-p\textbf{I} +\bigg[ {A}_0+\frac{1}{2} \hat{A}_1(t)(I_1-3)-
\hat{A}_2(t)\bigg]{\textbf{B}}+ \hat{A}_2(t){\textbf{B}}^{2},
\label{eq:26yy}
\end{align}
the superscript $\mathscr{R}_{\iso}$ referring to stress relaxation in an isotropic material.  

We may fix the pressure $p$ from  equation  
 (\ref{eq:25yy}) by the requirement that  $T^{\mathscr{R}_{\oi}}_{22}=0$ as
 \[ p= \left[{A}_0+\frac{1}{2} \hat{A}_1(t)
(\lambda^2-1)^2\lambda^{-2}
+\{ \hat{A}_2(t)+\hat{A}_6(t)\}(\lambda^{-2}-1)   \right]\lambda^{-2}.  \]
Equation (\ref{eq:25yy}) then    gives   the two non-zero pure shear tensions to be
\begin{align}
T^{\mathscr{R}_{\oi}}_{11}(\lambda, t)&= (\lambda^2-\lambda^{-2})
\left[{A}_0+\frac{1}{2} \hat{A}_1(t)
(\lambda^2-1)^2\lambda^{-2}
+ \hat{A}_2(t)(\lambda^2-1+\lambda^{-2}) \right]\nonumber\\
&\quad \mbox{}+ (\lambda^2-1) \left[\hat{A}_4(t)\lambda^2+\hat{A}_6(t)\lambda^{-4}\right], 
\label{eq:27yy} \\
T^{\mathscr{R}_{\oi}}_{33}(\lambda, t)&= (1-\lambda^{-2})
\left[{A}_0+\frac{1}{2} \hat{A}_1(t)
(\lambda^2-1)^2\lambda^{-2}
+ \left\{\hat{A}_2(t)+ \hat{A}_6(t)\right\}\lambda^{-2}    \right],
\label{eq:28yy}
\end{align}
with $T_{11}^{\mathscr{R}_{\oi}}(\lambda, t)$, $T_{33}^{\mathscr{R}_{\oi}}(\lambda, t)$ vanishing for $t\leq t_0^{\phantom{*}}$.

In (\ref{eq:27yy}) and (\ref{eq:28yy}), ${A}_0$ is a material constant and  $  \hat{A}_l(t)$,  where
$l\in \{1,2, 4, 6\}$, are material functions  which vanish for $t\leq t_0^{\phantom{*}}$ and are continuous for all $t$. 

If the material is now strained beyond the value $\lambda_{\C\_1}$ of stretch,  path $C$  continues onto path $A'$ as shown in Figure \ref{fig:tcreepcycle}. In the present model we assume that stress relaxation, given by equation (\ref{eq:25yy}), continues to evolve with time on the primary loading path $A'$,  i.e. path $P_2^{\phantom{*}}P_3^{\phantom{*}}$.  In straining the material beyond point $P_2^{\phantom{*}}$ to a point $P_3^{\phantom{*}}$ as shown in Figure \ref{fig:tcreepcycle} a new maximum stretch value $\lambda_{\C\_2}$ is imposed. 

For multiple stress-strain cycles, shown in Figure \ref{fig:tcreepcycle}, the functions $A_l(t)$  become
\begin{equation}
A_l(t)=\left\{\begin{array}{llll}
   \hat{A}_{l,1,0}(t)                                      & \textrm{primary loading}, & t_0^{\phantom{*}}\leq t\leq t_1^{\phantom{*}}, &  \textrm{path}\;\; P_0^{\phantom{*}}P_1^{\phantom{*}}\\[2mm]
\hat{A}_{l, {1,1}}(t)& \textrm{unloading}, & t_1^{\phantom{*}}\leq t\leq t_1^*, &  \textrm{path}\;\; P_1^{\phantom{*}}P_1^*\\[2mm]
\hat{A}_{l, {1,2}}(t)& \textrm{reloading}, & t^{*}_1\leq t\leq t_2^{\phantom{*}}, &  \textrm{path}\;\; P_1^{*}P_2^{\phantom{*}}\\[2mm]
 \;\; \dots&\;\; \dots&\;\; \dots&\;\; \dots \\[2mm]
\hat{A}_{l, {2,0}}(t)& \textrm{primary loading}, & t_3^{\phantom{*}}\leq t\leq t_4^{\phantom{*}}, &  \textrm{path}\;\; P_3^{\phantom{*}}P_4^{\phantom{*}}\\[2mm]
 \;\; \dots&\;\; \dots&\;\; \dots&\;\; \dots 
\end{array}\right.
\label{eq:29yy}
\end{equation}
in which $\hat{A}_{l, n,\omega}(t)$ are  continuous functions of time, with $n$ counting the number of cycles and $\omega$ being defined by equation (\ref{eq:24yy}). Note the occurrence of the functions $\hat{A}_{l, n,0}(t)$  because of the primary loading paths.

Employing equation (\ref{eq:29yy}), equation (\ref{eq:25yy})   becomes,
\begin{align}
\textbf{T}^{\mathscr{R}_{\oi}}(\lambda,t) & =-p\textbf{I} +\bigg[ {A}_0+\frac{1}{2} {A}_1(t)(I_1-3)-
{A}_2(t)\bigg]{\textbf{B}}+ {A}_2(t){\textbf{B}}^{2}\nonumber\\
&\;\;+{A}_4(t)(I_4-1)\bm{a}\otimes \bm{a}+{A}_6(t)(I_6-1)\bm{b}\otimes \bm{b},
\label{eq:30yy}
\end{align}
for  $ t>t_0^{\phantom{*}}$. The first line of (\ref{eq:30yy}) is the isotropic relaxation stress $\textbf{T}^{\mathscr{R}_{\iso}}(\lambda,t)$ as given by equation (\ref{eq:26yy}).

The total Cauchy stress for an orthotropic relaxing stress-softening material is then given by,

\begin{equation}
\mathbf{T}^{\oi}=\left\{\begin{array}{llll}
\zeta_{1,0}(\lambda)\{\textbf{T}^{\mathscr{E_{\iso}}}(\lambda)+\textbf{T}^{\mathscr{R_{\iso}}}(\lambda,t)\},& \textrm{primary loading}, & t_0^{\phantom{*}}\leq t\leq t_1^{\phantom{*}}, & \textrm{path}\;\; P_0^{\phantom{*}}P_1^{\phantom{*}}\\[2mm]                 
\zeta_{1,1}(\lambda)\{\textbf{T}^{\mathscr{E}_{\oi}}(\lambda) + \textbf{T}^{\mathscr{R}_{\oi}}(\lambda, t)\},& \textrm{unloading}, & t_1\leq t\leq t_1^{*}, &  \textrm{path}\;\; P_1P_1^*\\[2mm]
\zeta_{1,2}(\lambda)\{\textbf{T}^{\mathscr{E}_{\oi}}(\lambda) + \textbf{T}^{\mathscr{R}_{\oi}}(\lambda, t)\},& \textrm{reloading}, & t^{*}_1\leq t\leq t_2, &  \textrm{path}\;\; P_1^{*}P_2\\[2mm]
 \;\; \dots&\;\; \dots&\;\; \dots& \;\; \dots
\end{array}\right. 
 \label{eq:31yy}
 \end{equation}
where $\textbf{T}^{\mathscr{E}_{\oi}}(\lambda)$ is the orthotropic elastic stress (\ref{eq:22yy}) with $\textbf{T}^{\mathscr{E_{\iso}}}(\lambda)$ and $\textbf{T}^{\mathscr{R_{\iso}}}(\lambda,t)$ being defined by equations (\ref{eq:3yy}) and (\ref{eq:26yy}), respectively.

The total stress (\ref{eq:31yy}) falls to zero in $t>t_0^{\phantom{*}}$ and so we must have
$T^{\mathscr{R}_{\oi}}_{11}<0$ for $t>t_0^{\phantom{*}}$, implying that   $T^{\mathscr{R}_{\oi}}_{11}<0$ for $\lambda>1$.
Each of the quantities ${A}_0$, $A_1(t)$, $A_2(t)$, $A_4(t)$ and $A_6(t)$ occurring in equation (\ref{eq:30yy}) has positive coefficient for $\lambda>1$ and so at least one of them must be negative to maintain the requirement  $T^{\mathscr{R}_{\oi}}_{11}<0$ for $\lambda>1$.

In the literature on stress-relaxation we have been unable to identify an orthotropic version of the Bernstein \textit{et al.} \cite{bernstein} model.

\section{Orthotropic residual strain} 
\label{sec:orthresidual}
In this paper we assume minimal residual strain between the unloading paths during each cycle, i.e. in Figure \ref{fig:creepcycle2} we assume negligible separation between points $P_1^*$ and $P_2^*$, this observation being consistent with the experimental data of Figures \ref{fig:1ida2bw} and \ref{fig:1ida3bw} below.

For cyclic loading to multiple stress-strain cycles we employ a version of the  residual strain model developed by Rickaby and Scott \cite{rickaby2}:
\begin{equation}
\textbf{T}^{\mathscr{C}_{\oi}}(\lambda, t)=-p \textbf{I}+\left\{d(\lambda_{\C\_n})\left[\lambda_\chain-1\right]^{-1}\right\} \mathbf{B}, 
\label{eq:32yy}
\end{equation}
for $t > t_1^{\phantom{*}}$ and $\lambda>1$, with  $\textbf{T}^{\mathscr{C}_{\oi}}(\lambda,t)$ vanishing for $t\leq t_1^{\phantom{*}}$. In equation (\ref{eq:32yy}), $d(\lambda_{\C\_n})$ are material constants.  
The superscript $\mathscr{C}_{\oi}$ refers to residual strain in an orthotropic material.

For an orthotropic material the stretch of a polymer chain, denoted by $\lambda_\chain$, is given by:

\[
\lambda_\chain =  \frac{r_\chain}{r_0} =   \frac{\sqrt{I_4a^2+I_6b^2 +\left[I_1-I_4-I_6\right]c^2}}{\sqrt{a^2+b^2+c^2}} = \sqrt{N}\gamma,
\]
where $\gamma$ is given by equation (\ref{eq:gamma}).  Then equation (\ref{eq:32yy}) becomes
\begin{equation}
\textbf{T}^{\mathscr{C}_{\oi}}(\lambda, t)=-p \textbf{I}+\left\{d(\lambda_{\C\_n})\left[\sqrt{N}\gamma-1\right]^{-1}\right\} \mathbf{B}. 
\label{eq:34yy}
\end{equation} 

The total Cauchy stress for an orthotropic stress-softening relaxing material is now modelled by,

\begin{equation}
\mathbf{T}^{\oi}=\left\{\begin{array}{llll}
\zeta_{1,0}(\lambda)\textbf{T}^{\mathscr{E_{\iso}}+\mathscr{R_{\iso}}}(\lambda,t),& \textrm{primary loading}, & t_0^{\phantom{*}}\leq t\leq t_1^{\phantom{*}}, & \textrm{path}\;\; P_0^{\phantom{*}}P_1^{\phantom{*}}\\[2mm]                 
\zeta_{1,1}(\lambda)\textbf{T}^{{\mathscr{E}_{\oi}}+{\mathscr{R}_{\oi}}+{\mathscr{C}_{\oi}} }(\lambda, t),& \textrm{unloading}, & t_1^{\phantom{*}}\leq t\leq t_1^{*}, &  \textrm{path}\;\; P_1^{\phantom{*}}P_1^*\\[2mm]
\zeta_{1,2}(\lambda)\textbf{T}^{{\mathscr{E}_{\oi}}+{\mathscr{R}_{\oi}}+{\mathscr{C}_{\oi}} }(\lambda, t),& \textrm{reloading}, & t^{*}_1\leq t\leq t_2^{\phantom{*}}, &  \textrm{path}\;\; P_1^{*}P_2^{\phantom{*}}\\[2mm]
 \;\; \dots&\;\; \dots&\;\; \dots&\;\; \dots
 \end{array}\right. 
 \label{eq:35yy}
 \end{equation}
in which for notational convenience we have defined the stresses
\begin{align}
\textbf{T}^{\mathscr{E}_{\iso} + \mathscr{R}_{\iso}}(\lambda, t)  =&\, \textbf{T}^{\mathscr{E}_{\iso}}(\lambda) + \textbf{T}^{\mathscr{R}_{\iso}}(\lambda, t), \nonumber \\[4pt]
  \textbf{T}^{{\mathscr{E}_{\oi}}+{\mathscr{R}_{\oi}}+{\mathscr{C}_{\oi}} }(\lambda, t)  =&\, \textbf{T}^{\mathscr{E}_{\oi}}(\lambda) + \textbf{T}^{\mathscr{R}_{\oi}}(\lambda, t) + \textbf{T}^{\mathscr{C}_{\oi}}(\lambda, t), 
\label{eq:36yy}
\end{align}
where $\textbf{T}^{\mathscr{E}_{\iso}}(\lambda)$,
$\textbf{T}^{\mathscr{R}_{\iso}}(\lambda,t)$, $\textbf{T}^{\mathscr{E}_{\oi}}(\lambda)$, $\textbf{T}^{\mathscr{R}_{\oi}}(\lambda,t)$ and $ \textbf{T}^{\mathscr{C}_{\oi}}(\lambda, t)$ are given by equations 
(\ref{eq:3yy}), (\ref{eq:26yy}), (\ref{eq:22yy}), (\ref{eq:30yy}) and (\ref{eq:34yy}), respectively.

\section{Softening on the subsequent primary loading paths} 
\label{sec:shearsofteningA}

Referring to Figure \ref{fig:tcreepcycle}, if the material had not been unloaded from point $P_1$, but instead loading had continued to greater stretches, then the resulting primary loading path would be the dashed path $\bar{A}$ marked in this figure. From the experimental data of Diani \textit{et al.} \cite[Figure 1]{dianib} it is observed that the new primary loading paths, namely path $A'$ and $A''$ of Figure \ref{fig:tcreepcycle},  tend towards, or return to, the primary loading path $\bar{A}$. To account for this feature, Rickaby and Scott \cite{rickaby3} introduced the following damage function:
\begin{equation}
\zeta_{n,0}^{\phantom{*}}(\lambda)=1-\frac{1}{r_{\C\_n}}\left\{\tanh\left(\frac{\lambda_{\C\_n}-{\lambda}}{ b_3}\right)\right\}^{{1}/{\vartheta_3}}, \quad \where \quad  \lambda_{\C\_(n-1)} \leq  \lambda\leq\lambda_{\C\_n},
\label{eq:37yy}
\end{equation}
with $b_3$,  $\vartheta_3$, $r_{\C\_n}$ being material constants chosen to satisfy the condition that $\zeta_{n,0}(\lambda)> 0 $ on the subsequent primary loading paths, $n$ counting the number of cycles.

The new primary loading paths may be modelled by combining $\zeta_{n,0}^{\phantom{*}}(\lambda)$ with the total stress for the orthotropic material $\textbf{T}^{\oi}(\lambda, t)$ on the primary loading path, which is obtained by summing together all the different stress components:
\begin{equation*}
\textbf{T}^{\oi}(\lambda, t)  = \zeta_{n,0}^{\phantom{*}}(\lambda)\textbf{T}^{{\mathscr{E}_{\oi}}+{\mathscr{R}_{\oi}}+{\mathscr{C}_{\oi}} }(\lambda, t),
\end{equation*}
where $\textbf{T}^{{\mathscr{E}_{\oi}}+{\mathscr{R}_{\oi}}+{\mathscr{C}_{\oi}} }(\lambda, t)$ is given by equation (\ref{eq:36yy})$_2$.

\section{Constitutive model} 
\label{sec:constitutive}
From equations (\ref{eq:35yy}) and (\ref{eq:37yy}) the general constitutive stress-softening model for cyclic loading to multiple stress-strain cycles is given by:
\begin{equation}
\mathbf{T}^{\oi}=\left\{\begin{array}{llll}
\zeta_{1,0}(\lambda)\textbf{T}^{\mathscr{E_{\iso}}+\mathscr{R_{\iso}}}(\lambda,t),& \textrm{primary loading}, & t_0^{\phantom{*}}\leq t\leq t_1^{\phantom{*}}, & \textrm{path}\;\; P_0^{\phantom{*}}P_1^{\phantom{*}}\\[2mm]                 
\zeta_{1,1}(\lambda)\textbf{T}^{{\mathscr{E}_{\oi}}+{\mathscr{R}_{\oi}}+{\mathscr{C}_{\oi}} }(\lambda, t),& \textrm{unloading}, & t_1^{\phantom{*}}\leq t\leq t_1^{*}, &  \textrm{path}\;\; P_1^{\phantom{*}}P_1^*\\[2mm]
\zeta_{1,2}(\lambda)\textbf{T}^{{\mathscr{E}_{\oi}}+{\mathscr{R}_{\oi}}+{\mathscr{C}_{\oi}} }(\lambda, t),& \textrm{reloading}, & t^{*}_1\leq t\leq t_2^{\phantom{*}}, &  \textrm{path}\;\; P_1^{*}P_2^{\phantom{*}}\\[2mm]
 \;\; \dots&\;\; \dots&\;\; \dots&\;\; \dots\\[2mm]
{\zeta_{2,0}(\lambda)}\textbf{T}^{{\mathscr{E}_{\oi}}+{\mathscr{R}_{\oi}}+{\mathscr{C}_{\oi}} }(\lambda, t),& \textrm{primary loading}, & t_3^{\phantom{*}}\leq t\leq t_4^{\phantom{*}}, &  \textrm{path}\;\; P_3^{\phantom{*}}P_4^{\phantom{*}}\\[2mm]
 \;\; \dots&\;\; \dots&\;\; \dots&\;\; \dots
 \end{array}\right. 
  \label{eq:38yy}
 \end{equation}
where once again the stresses $\textbf{T}^{\mathscr{E_{\iso}}+\mathscr{R_{\iso}}}(\lambda,t)$ and $\textbf{T}^{{\mathscr{E}_{\oi}}+{\mathscr{R}_{\oi}}+{\mathscr{C}_{\oi}} }(\lambda,t)$, defined by (\ref{eq:36yy}), are employed for notational convenience.

On substituting the individual stress components given by equations (\ref{eq:3yy}), (\ref{eq:26yy}), (\ref{eq:22yy}), (\ref{eq:30yy}) and (\ref{eq:34yy}) into equation (\ref{eq:38yy}) we obtain the following model for an orthotropic material during repeated unloading and reloading, displaying:   softening, hysteresis, stress relaxation,  residual strain
\begin{align}
\textbf{T} =&\,\left[1-\frac{1}{r_\omega}\left\{\tanh\left(\frac{W_{\C\_n}-{W}}{\mu b_\omega}\right)\right\}^{{1}/{\vartheta_\omega}}\right]\times\nonumber\\
&\times\Bigg\{ -p\mathbf{I} + 
\mu\frac{1}{1+\alpha_1^2+\alpha_2^2} \gamma^{-1}\beta \bigg\{\alpha^2_2\mathbf{B} +
(1-\alpha^2_2)\bm{a}\otimes\bm{a}+(\alpha_1^2-\alpha^2_2)\bm{b}\otimes\bm{b}\bigg\} - h_4 \bm{a}\otimes\bm{a} -  h_6\bm{b}\otimes\bm{b} \nonumber \\
&\qquad+\bigg[ {A}_0+\frac{1}{2} {A}_1(t)(I_1-3)-
{A}_2(t)\bigg]{\textbf{B}}+ {A}_2(t){\textbf{B}}^{2}\nonumber\\
&\qquad+{A}_4(t)(I_4-1)\bm{a}\otimes \bm{a}+{A}_6(t)(I_6-1)\bm{b}\otimes \bm{b}\nonumber\\
&\qquad+d(\lambda_{\C\_n})\left[\sqrt{N}\gamma-1\right]^{-1}  \textbf{B}
\Bigg\}.
\label{eq:39yy}
\end{align}
In modelling the Mullins effect we have used the engineering (nominal) stress component 
\[
T_{E11}=\lambda^{-1}T_{11}
\]
for ease of comparison with experimental data.

\section{Comparison with experimental data} 
\label{sec:experimental}

\begin{figure}[h] 
\centerline{
\includegraphics[width=18cm,height=10cm]{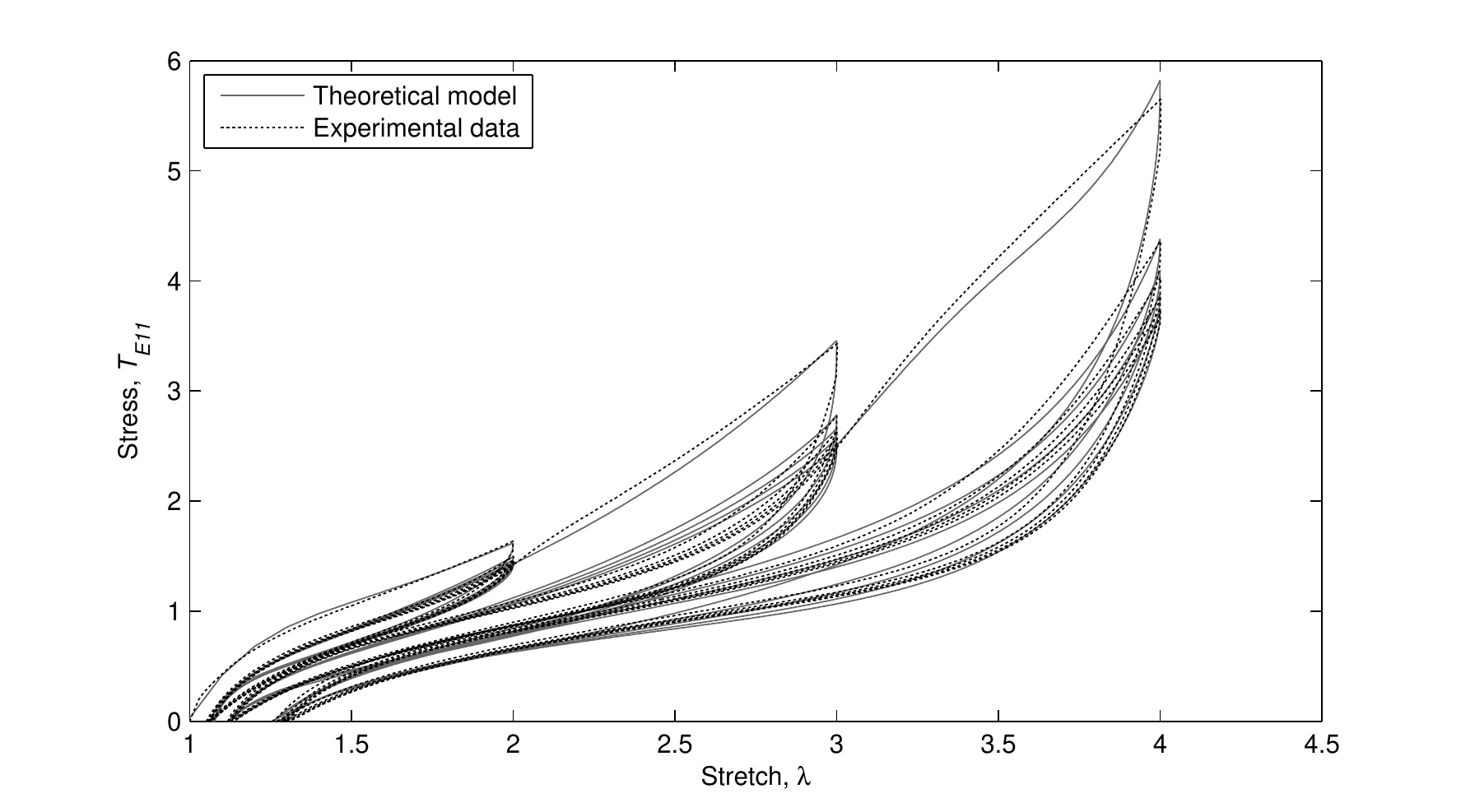}}
\caption{Comparison with experimental data of Raoult \textit{et al.} \cite{raoult}, carbon black reinforced natural rubber with 43 phr of carbon black, material sample A.}
\label{fig:1ida2bw}
\end{figure}

\begin{figure}[h] 
\centerline{
\includegraphics[width=18cm,height=10cm]{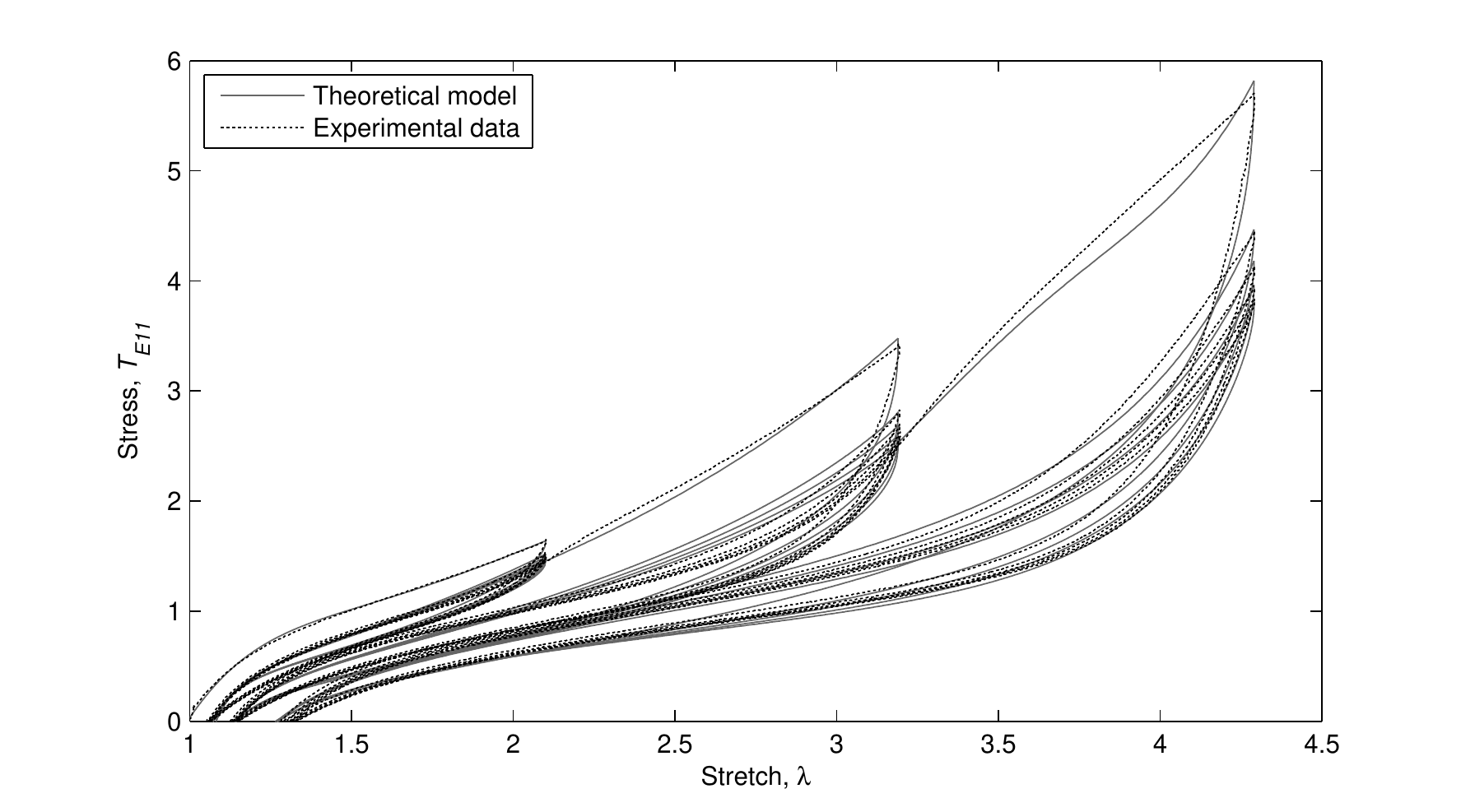}}
\caption{Comparison with experimental data of Raoult \textit{et al.} \cite{raoult}, carbon black reinforced natural rubber with 43 phr of carbon black, material sample B.}
\label{fig:1ida3bw}
\end{figure}

Figures \ref{fig:1ida2bw} and \ref{fig:1ida3bw} provide a comparison of the orthotropic constitutive model we have developed with experimental data. The experimental
data came courtesy of Trelleborg and PSA Peugeot Citro\"en, and was partly presented in the paper of Raoult \cite{raoult}. The experimental data is for two different material samples, A and B, though both samples are vulcanized natural rubber and contain the same filler concentration.

Figure \ref{fig:1ida2bw} has been obtained by using the following constants and functions:
\[
N=7.2, \quad \mu=0.710,  \quad \alpha_1^2=1.8,
\quad
{A}_0=-0.005, \quad A_{1,2,4,6}(t)=-0.006\log(0.5t),
\]
\[
r=\left\{ \begin{array}{clrr}
2.00\\
2.00\\
\end{array}\right.
   \quad
\alpha_2^2=\left\{ \begin{array}{clrr}
0.25\\
0.35\\
\end{array}\right.
\quad
\vartheta_\omega=\left\{\begin{array}{clrr}
0.40& \textrm{unloading},\\
0.70& \textrm{loading}.\\
\end{array}\right.
\]
For $\lambda_{\C\_1}=2.0$
\[
\zeta_{1,0}(\lambda)=1+0.55{[ \tanhs(\lambda_{\C\_1}-\lambda)]^{3.5}},
\quad
d(\lambda_{\C\_1})=0.04,  
\quad
\mu b_\omega=\left\{\begin{array}{clrr}
\s1.10& \textrm{unloading},\\
\s4.00& \textrm{loading}.\\
\end{array}\right.
\]
For $\lambda_{\C\_2}=3.0$
\[
\zeta_{1,0}(\lambda)=1-0.35{[ \tanhs(\lambda_{\C\_2}-\lambda)]^{4}},
\quad
d(\lambda_{\C\_2})=0.07,
\quad
\mu b_\omega=\left\{\begin{array}{clrr}
\s1.15& \textrm{unloading},\\
\s4.00& \textrm{loading}.\\
\end{array}\right.
\]
For $\lambda_{\C\_3}=4.0$
\[
\zeta_{1,0}(\lambda)=1-0.95{[ \tanhs(\lambda_{\C\_3}-\lambda)]^{4}},
\quad
d(\lambda_{\C\_3})=0.15,  
\quad
\mu b_\omega=\left\{\begin{array}{clrr}
\s3.80& \textrm{unloading},\\
35.00& \textrm{loading}.\\
\end{array}\right.
\]
We see in Figure \ref{fig:1ida2bw} that the orthotropic model developed here provides a good fit with experimental~data.

Figure \ref{fig:1ida3bw} has been obtained by using the following constants and functions,
\[
N=7.2, \quad \mu=0.666, \quad \alpha_1^2=2.3, 
\quad
{A}_0=-0.005, \quad A_{1,2,4,6}(t)=-0.005\log(0.5t),
\]
\[
r=\left\{ \begin{array}{clrr}
2.00\\
2.00\\
\end{array}\right.
\quad
\alpha_2^2=\left\{ \begin{array}{clrr}
0.25\\
0.35\\
\end{array}\right.
\quad
\vartheta_\omega=\left\{\begin{array}{clrr}
0.40& \textrm{unloading},\\
0.70& \textrm{loading}.\\
\end{array}\right.
\]
For $\lambda_{\C\_1}=2.1$
\[
\zeta_{1,0}(\lambda)=1+0.55{[ \tanhs(\lambda_{\C\_1}-\lambda)]^{3.5}},
\quad
d(\lambda_{\C\_1})=0.04,
\quad
\mu b_\omega=\left\{ \begin{array}{clrr}
\s1.40& \textrm{unloading},\\
\s4.50& \textrm{loading}.\\
\end{array}\right.
\]
For $\lambda_{\C\_2}=3.2$
\[
\zeta_{2,0}(\lambda)=1-0.35{[ \tanhs(\lambda_{\C\_2}-\lambda)]^{5}},
\quad
d(\lambda_{\C\_2})=0.08, 
\quad
\mu b_\omega=\left\{\begin{array}{clrr}
\s1.20& \textrm{unloading},\\
\s4.50& \textrm{loading}.\\
\end{array}\right.
\]
For $\lambda_{\C\_3}=4.3$
\[
\zeta_{3,0}(\lambda)=1-0.95{[ \tanhs(\lambda_{\C\_3}-\lambda)]^{5}},
\quad
d(\lambda_{\C\_3})=0.17,  
\quad
\mu b_\omega=\left\{\begin{array}{clrr}
\s2.80& \textrm{unloading},\\
25.00& \textrm{loading}.\\
\end{array}\right.
\]
As can be seen from Figure \ref{fig:1ida3bw} the orthotropic model we have developed is shown to provide good agreement with experimental data.

The experimental data of material samples A and B as given in Figures \ref{fig:1ida2bw} and \ref{fig:1ida3bw}, respectively are very similar. For both material samples the stress at the start of unloading for cycle 1 is approximately 1.64 MPa; for material sample A the associated stretch needed to achieve this stress value is $\lambda_{\C\_1}=2$ and for material sample B the associated stretch is $\lambda_{\C\_1}=2.1$. The stress at the start of unloading for cycle 2 for both material samples A and B is approximately 3.40 MPa; for material sample A the associated stretch needed to achieve this stress value is $\lambda_{\C\_2}=3$ and for material sample B the associated stretch is $\lambda_{\C\_2}=3.19$. For material sample A the stress at the start of unloading for cycle 3 is approximately 5.60 MPa with associated stretch $\lambda_{\C\_3}=4$, and for material sample B the stress at the start of unloading for cycle 3 is approximately 5.70 MPa with associated stretch $\lambda_{\C\_3}=4.29$. For both material samples A and B the increase in stress at the start of unloading for cycles 1 and 2 are roughly comparable, with the increase in stress at the start of unloading for cycle 3 being greater. 

\section{Conclusions}
\label{sec:conclusion}
From Figures \ref{fig:1ida2bw} and \ref{fig:1ida3bw} it is seen that the orthotropic model provides an excellent fit with the experimental data. The close similarity between the two material samples presented in Figures \ref{fig:1ida2bw} and \ref{fig:1ida3bw} is captured in the model we have developed here by having different material constants only for $\mu$ and $b_\omega$. This demonstrates that once material parameters have been determined for a specific rubber vulcanizate then the model could be used to predict the behaviour of other rubber vulcinazates with a corresponding molecular structure.

To the best of our knowledge this is the first time that an orthotropic stress-softening and residual strain model has been combined with an orthotropic version of the Arruda-Boyce eight-chain constitutive equation in order to develop a model that is capable of representing the Mullins effect for an orthotropic, incompressible, hyperelastic material.

We see in Figures \ref{fig:1ida2bw} and \ref{fig:1ida3bw} that the curves occupy quite narrow bands along the $\lambda$-axis.  This shows that there is very little creep of residual strain present in the experimental data, thus justifying the omission of this effect from the present model.
The results presented in Figures \ref{fig:1ida2bw} and \ref{fig:1ida3bw} are by no means the only solutions that this model is capable of giving. By neglecting, or limiting the accuracy of, any of the modelled inelastic
terms, i.e.  selecting a single relaxation curve, there results  a simplified model with a reduced set of parameters. The generalized model developed here is shown to produce an accurate representation of the Mullins effect for a pure shear deformation. The model has been developed in such a way that any of the salient inelastic features, could be excluded and the integrity of the model would still be maintained.

Dorfmann and Pancheri  \cite{dorfmann2012} conducted a series of experiments to assess the degree of deformation-induced anisotropy in particle filled rubber. They observe that the deformation of rubber induces a change in the properties of the material, generating a preferred direction, that is, an initially isotropic material becomes anisotropic.  These observations are echoed by several authors, see for example Dargazany and Itskov \cite{dargazany} and Machado et al.\@ \cite{machado2014}.  Unfortunately,  for pure shear loading, no conclusions have yet been drawn in the literature as to the anisotropic form of the material after initial primary loading.

A further application of this model could be in the development of earthquake protective systems, through rubber seismic isolation flexible bearings. One of the most effective bearings is the lead-rubber bearing, see, for example, Dowrick \cite[pages 295-296 ]{dowrick}. It has been found experimentally that lead-rubber bearings deform in pure shear, see Islam \cite{islam},  with the rubber component exhibiting stress relaxation, hysteresis and residual strain, all of which can be modelled by means of the model developed here. 

\section*{Acknowledgements}
One of us (SRR) is grateful to  the University of East Anglia for the award of a PhD studentship. The authors thank Dr Ida Raoult, Dr Pierre Charrier, Trelleborg and PSA Peugeot Citro\"en for most kindly supplying  experimental data. Furthermore,  we would like to thank the reviewers for their constructive comments and suggestions.

\bibliographystyle{plain}
\bibliography{BIBLIOGRAPHYa}

\appendix

\section{Derivation of equation (\ref{eq:9yy})}
\label{sec:appendix}
The derivation of equation (\ref{eq:9yy}) is based upon the Cayley-Hamilton theorem for the  $3 \times 3$ tensor $\bf D$:
\begin{equation}
\bm{D}^3-\bm{D}^2\tr\bm{D}+\frac{1}{2}\bm{D}\left\{(\tr\bm{D})^2-\tr\bm{D}^2\right\}-\textbf{I}\det\bm{D}=\textbf{0},
\label{eq:a1}
\end{equation}
where $\textbf{0}$ is the $3 \times 3$ zero matrix.  Taking the trace of  (\ref{eq:a1}) gives 
\begin{equation*}
\det(\bm{D})=\frac{1}{6}(\tr\bm{D})^3-\frac{1}{2}\tr\bm{D}\tr\bm{D}^2+\frac{1}{3}\tr\bm{D}^3,
\end{equation*}
which may be combined with equation (\ref{eq:a1}) to give
\begin{equation}
\bm{D}^3-(\tr\bm{D})\bm{D}^2+\frac{1}{2}\left\{(\tr\bm{D})^2-\tr\bm{D}^2\right\}\bm{D}-\left\{\frac{1}{6}(\tr\bm{D})^3-\frac{1}{2}\tr\bm{D}\tr\bm{D}^2+\frac{1}{3}\tr\bm{D}^3\right\}\textbf{I}=\textbf{0}.
\label{eq:a2}
\end{equation}

Following  Rivlin \cite{rivlin1955}, we  set $\bm{D}=\bm{A}+\bm{B}$, and $\bm{D}=\bm{A}-\bm{B}$, in turn, and subtract the two resulting equations to give
\begin{align}
&\,\bm{A}\bm{B}\bm{A}+\bm{B}\bm{A}^2+\bm{A}^2\bm{B}-\left(\bm{A}\bm{B}+\bm{B}\bm{A}\right)\tr\bm{A}\nonumber\\
&\;\;-\, \bm{A}^2\tr\bm{B}+\bm{A}\Big\{\tr\bm{A}\tr\bm{B}-\tr\bm{B}\bm{A}\Big\}+\frac{1}{2}\bm{B}\Big\{(\tr\bm{A})^2-\tr\bm{A}^2\Big\}
\nonumber\\
&\;\;\;\;+\, \textbf{I}\bigg(\tr\bm{A}\tr\bm{B}\bm{A}-\tr\bm{A}^2\bm{B}-\frac{1}{2}\tr\bm{B}\Big\{(\tr\bm{A})^2-\tr\bm{A}^2\Big\}\bigg)=\textbf{0},
\label{eq:a7}
\end{align}
where the Cayley-Hamilton theorem, in the form (\ref{eq:a2}), for $\bf B$ has been used.

Replacing $\bf A$ by $\textbf{C}$, the right Cauchy-Green strain tensor, and  $\bm{B}$ by  $\textbf{u}\otimes\textbf{v}$, in equation (\ref{eq:a7}) leads to the relation
\begin{align}
&\textbf{C}(\textbf{u}\otimes\textbf{v})\textbf{C}+(\textbf{u}\otimes\textbf{v})\textbf{C}^2+\textbf{C}^2(\textbf{u}\otimes\textbf{v})-\Big\{\textbf{C}(\textbf{u}\otimes\textbf{v})+(\textbf{u}\otimes\textbf{v})\textbf{C}\Big\}\tr\textbf{C}-\textbf{C}^2(\textbf{u}\cdot\textbf{v})\nonumber\\
&\;\;+\,2(\textbf{u}\cdot\textbf{v})\Big\{\textbf{C}\tr\textbf{C}-\textbf{C}^2\Big\}+\frac{1}{2}((\textbf{u}\otimes\textbf{v})-\textbf{I}(\textbf{u}\cdot\textbf{v}))\Big\{(\tr\textbf{C})^2-\tr\textbf{C}^2\Big\}=\textbf{0}.
\label{eq:a8}
\end{align}

Following  Spencer \cite{Spencer2009}, we  pre-multiply equation (\ref{eq:a8}) by $\textbf{u}$ and post-multiply by $\textbf{v}$, to derive the following identity relating the  ten invariants defined by equations (\ref{eq:2yy}) and (\ref{eq:8yy}): 
\begin{align}
&(\textbf{u}\cdot(\textbf{C}\textbf{u}))(\textbf{v}\cdot(\textbf{C}\textbf{v}))
+\textbf{v}\cdot(\textbf{C}^2\textbf{v})
+\textbf{u}\cdot(\textbf{C}^2\textbf{u})
-\Big\{\textbf{u}\cdot(\textbf{C}\textbf{u})+\textbf{v}\cdot(\textbf{C}\textbf{v})\Big\}\tr\textbf{C}-(\textbf{u}\cdot(\textbf{C}\textbf{v}))^2\nonumber\\
&\;\;
+\, 2(\textbf{u}\cdot\textbf{v})\Big\{(\textbf{u}\cdot(\textbf{C}\textbf{v}))\tr\textbf{C}
-(\textbf{u}\cdot(\textbf{C}^2\textbf{v}))\Big\}+\frac{1}{2}(1-(\textbf{u}\cdot\textbf{v})^2)\Big\{(\tr\textbf{C})^2-\tr\textbf{C}^2\Big\}=0.
\label{eq:a9}
\end{align}
Equation (\ref{eq:9yy}) is obtained by rearranging 
(\ref{eq:a9}) and using the identity,
\[
(\textbf{u} \times \textbf{v})\cdot(\textbf{u} \times \textbf{v})=1-(\textbf{u}\cdot\textbf{v})^2.
\]

If  $\bm{u}$ and $\bm{v}$ are no longer unit vectors, equation (\ref{eq:9yy}) is replaced by the identity
\begin{align}
&\frac{1}{2}(\textbf{u} \times \textbf{v})\cdot(\textbf{u} \times \textbf{v})\left\{(\tr\textbf{C})^2-\tr\textbf{C}^2\right\}
+2(\bm{u}\cdot\bm{v})\left\{(\bm{u}\cdot(\textbf{C}\bm{v}))\tr\textbf{C}-\bm{u}\cdot(\textbf{C}^2\bm{v})\right\}\nonumber\\
&\;\; -\, \left\{(\bm{u}\cdot(\textbf{C}\bm{u}))(\bm{v}\cdot\bm{v})+(\bm{v}\cdot(\textbf{C}\bm{v}))(\bm{u}\cdot\bm{u})\right\}\tr\textbf{C}+(\bm{u}\cdot(\textbf{C}\bm{u}))(\bm{v}\cdot(\textbf{C}\bm{v}))-(\bm{u}\cdot(\textbf{C}\bm{v}))^2\nonumber\\
&\;\;\;\;
+\, \bm{u}\cdot(\textbf{C}^2\bm{u})(\bm{v}\cdot\bm{v})+\bm{v}\cdot(\textbf{C}^2\bm{v})(\bm{u}\cdot\bm{u})=
0.  
\end{align}

\end{document}